\documentclass{iaus}
\usepackage{graphicx}

\title[Structure of Stellar Clusters]{Internal Structure
of Stellar Clusters: \\ Geometry of Star Formation}

\author[Emilio J. Alfaro \& N\'estor S\'anchez]
{Emilio J. Alfaro \& N\'estor S\'anchez}

\affiliation{Instituto de Astrof{\'i}sica de Andaluc{\'i}a, CSIC,
Apdo. 3004, E-18080, Granada, Spain}

\pubyear{2011}
\volume{270}
\pagerange{001--008}
\jname{Computational Star Formation}
\editors{J. Alves, B. Elmegreen, J. Girart \& V. Trimble, eds.}

\begin{document}

\maketitle

\begin{abstract}
The study of the internal structure of star clusters provides
important clues concerning their formation mechanism and dynamical 
evolution. There are both observational and numerical evidences
indicating that open clusters evolve from an initial clumpy
structure, presumably a direct consequence of the formation
in a fractal medium, toward a centrally condensed state.
This simple picture has, however, several drawbacks. There can be 
very young clusters exhibiting radial patterns maybe reflecting the 
early effect of gravity on primordial gas. There can be also very 
evolved cluster showing fractal patterns that either have survived 
through time or have been generated subsequently by some (unknown) 
mechanism. Additionally, the fractal structure of some open clusters
is much clumpier than the average structure of the interstellar
medium in the Milky Way, although in principle a very similar
structure should be expected. Here we summarize and discuss
observational and numerical results concerning this subject.
\end{abstract}

\firstsection

\section{Introduction}

Most of visible matter in the universe is condensed into
stars, with densities more than 30 orders of magnitude higher
than the average density of the universe and more than 20 orders
of magnitude higher than the densities of the interstellar clouds
in which they form (Larson 2007). Thus, the fundamental question 
is not how baryons end up as stars, but how some of them form
stars and others remain as hot, low-density interstellar gas.
This enigma lies at the core of a predictive theory of star
formation, one of the main goals of modern astronomy. We are
still far away from a global solution to this complex problem,
whose answer depends very much on the existence of a
well-structured and complete set of empirical data, as well
as on the building of reliable and precise simulation tools. 

Nowadays, it is widely accepted that stars form in highly
hierarchical stellar systems that mimic, in some way, the
stepped structure of the interstellar medium (ISM) or, at
least, the morphology of the densest regions.
This hierarchical pattern, both spatial and temporal, presents
singular condensations, the stellar clusters, whose main physical 
characteristics make them reliable tracers of the star forming
processes in galaxies. Hierarchical structure extends from star
complexes (or large portions of spiral arms in flocculent galaxies)
through embedded clusters to individual young stars inside those
embedded clusters. The cluster scale is the best metric to measure
and analyze the whole spatial range in the formation of stellar
systems. 

The study of star forming regions in the infrared range led to the
conclusion that most stars, if not all, born grouped in clusters
(Lada \& Lada 2003). However, after ten million years, the
fraction of stars in clusters is reduced to 10\% and this figure
tends to decrease with age. This simple temporal pattern suggests
that a few million years after their birth star clusters suffer a
high mortality rate. The survivors evolve then under different
destruction processes and are gradually eroded until diluted into
the galactic field.

The initial conditions, that is, the properties of the
cold and dark clouds that eventually form stars are poorly
known (Bergin and Tafalla 2007). 
A few years ago, we started a project aimed to characterize the
geometry of the ISM. This information would provide important
clues on the physical processes developing and maintaining the
internal structure of the clouds. Since the pioneering work by
Larson (1981), turbulence is considered the best candidate to do this
job. It appears that the distribution of gas and dust in these clouds 
determines the initial conditions of a newborn cluster because 
star formation follows the patterns defined by the densest regions 
(Bonnell et al. 2003). Thus, the fractal (self-similar) distribution 
of the gas in molecular cloud complexes may account for the hierarchical 
structure observed in some open clusters (Elmegreen 2010). However, 
observations show that the morphologies of young clusters show a 
wide variety, from hierarchical to centrally condensed ones, often 
being elongated or surrounded by a low-density stellar halo (see, 
for example, Ma\'{\i}z-Apell\'aniz 2001; Hartmann 2002; and Caballero
2008). The reasons for this heterogeneity of shapes are still poorly
understood.

Here we review and discuss the formation of star clusters and
their evolution from a geometric point of view.
In particular, our framework is
determined by several questions: What do physical mechanisms
control and shape the internal structure of fertile gas clouds?
What is the influence of the geometric structure of a star-forming
cloud on the internal spatial structure of the stellar population
formed from it? How does it evolve with time? Because of the
hierarchical and self-similar nature involved, fractal geometry
appears to be a good descriptor for these physical structures.
Thus, the design and development of mathematical tools for
determining the fractal dimension of 3D gas clouds and point-like
object distributions are also discussed and evaluated in this work.

\section{Fractal dimension of the interstellar medium}

Gas and dust in the Galaxy
are organized into irregular structures in a hierarchical
and approximately self-similar manner. This means that interstellar
clouds can be well described or characterized as fractal structures
(Mandelbrot 1983). Many tools can be used to characterize the
complexity of these structures (see Elmegreen \& Scalo 2004)
but to measure the fractal dimension seems
particularly appropriate when dealing with nearly fractal
systems. The measurement of the dimension of the projected
boundaries $D_{per}$ is the most used method to characterize
the fractal properties of interstellar clouds, but there is
a wide variation in the estimated values. A summary of results
that can be accessed via NASA's ADS service is shown in
Table~\ref{tabDper}.
\begin{table}
\begin{center}
\caption{Summary of perimeter dimensions
in molecular clouds in the Galaxy}
\label{tabDper}
{\scriptsize
\begin{tabular}{|c|l|l|}\hline 
Ref. & $D_{per}$ Ê& Region/Map \\
\hline
(1)  & $1.40$      & Extinction maps of dark clouds \\
(2)  & $1.12-1.40$ & Dust emission maps of cirrus clouds \\
(3)  & $1.17-1.30$ & Infrared intensity and column density
                     maps of several molecular clouds \\
(4)  & $1.36$      & Molecular emission maps (Taurus complex) \\
(5)  & $1.38-1.52$ & Visual extinction maps (Chamaeleon complex) \\
(6)  & $1.51$      & Molecular emission map (Taurus complex) \\
(7)  & $1.23-1.54$ & HI maps of high-velocity clouds and
                     infrared maps of cirrus clouds \\
(8)  & $1.34-1.40$ & Molecular emission maps of clouds 
                     in the antigalactic center \\
(9)  & $1.31-1.35$ & Molecular emission maps 
                     (Ophiuchus, Perseus, and Orion clouds) \\
(10) & $1.50-1.53$ & Molecular emission maps of clouds in the outer Galaxy \\
\hline
\end{tabular}}
\end{center}
\vspace{1mm}
\scriptsize{
{\it Reference index:}
(1) Beech (1987); 
(2) Bazell \& Desert (1988); 
(3) Dickman et al. (1990); 
(4) Falgarone et al. (1991); 
(5) Hetem \& Lepine (1993); 
(6) Stutzki (1993); 
(7) Vogelaar \& Wakker (1994); 
(8) Lee (2004); 
(9) S\'anchez et al. (2007a); 
(10) Lee et al. (2008).}
\end{table}

In general, observed values are spread over the range $1.1
\lesssim D_{per} \lesssim 1.5$. It is not clear, however,
whether the different values seen in Table~\ref{tabDper}
represent real variations or they are consequence of
different data quality and/or analysis techniques.
For example, it is well known that the obtained results may be
affected by factors such as image resolution and/or signal-to-noise
ratio (Dickman et al. 1990; Vogelaar \& Wakker 1994;
Lee 2004; S\'anchez et al. 2005, 2007a). Note, for example,
that for CO emission maps of the same region in the Taurus
molecular complex Falgarone et al. (1991) obtained $D_{per}=
1.36$ whereas Stutzki (1993) found $D_{per}=1.51$ on a different
set of data. Despite those results,
the general ``belief" is that the fractal
dimension of the projected boundaries of interstellar clouds is
roughly a constant throughout the Galaxy, with $D_{per} \simeq
1.3-1.4$ (Bergin \& Tafalla 2007). This constancy in $D_{per}$
would be a natural consequence of a universal picture in which
interstellar turbulence is driven by the same physical mechanisms 
everywhere (Elmegreen \& Scalo 2004).
In order to get reliable clues about the ISM 
structure, it is important that any analysis technique 
is applied systematically on homogeneous data sets.
S\'anchez et al. (2007a) used several maps of different
regions (Ophiuchus, Perseus, and Orion molecular clouds)
in different emission lines and calculated $D_{per}$ by
using an algorithm previously calibrated on simulated fractals 
(S\'anchez et al. 2005). In this case the range of obtained values 
decreased notoriously to $1.31 \lesssim D_{per} \lesssim 1.35$
(reference~$9$ in Table~\ref{tabDper}).

But what is the corresponding value of the fractal dimension of
interstellar clouds in the three-dimensional space, $D_f$?
It has been traditionally assumed that $D_f = D_{per}+1 \simeq
2.3-2.4$ (Beech 1992). S\'anchez et al. (2005) used simulated
fractal clouds to study the relationship between $D_{per}$
and $D_f$ and showed this assumption is not correct. Their
main result (see their Figure~$8$) indicate that if the perimeter dimension
is around $D_{per} \simeq 1.31-1.35$ (S\'anchez et al. 2007a) then
the 3D fractal dimension should be in the range $D_f \sim 2.6-2.8$.
This dimension is clearly higher than the value $D_f \sim 2.3$
that is usually assumed in the literature for interstellar clouds
in the Galaxy (Bergin \& Tafalla 2007).

\section{Fractal dimension of young stellar clusters}

The distribution of stars and star-forming regions also exhibits a
spatial hierarchy from large star complexes to individual clusters
(Efremov 1995; de la Fuente Marcos \& de la Fuente Marcos 2006,
2009; Elias et al. 2009; Elmegreen 2010).
This hierarchical structure is presumably a direct consequence 
of the fact that stars are formed in a medium with an underlying 
fractal structure (previous Section). If this were the case, then
it is reasonable to assume that the fractal dimension of the 
distribution of new-born stars should be nearly the same as 
that of the molecular clouds from which they are formed.

\begin{figure}[thb]
\begin{center}
\includegraphics[width=3.4in]{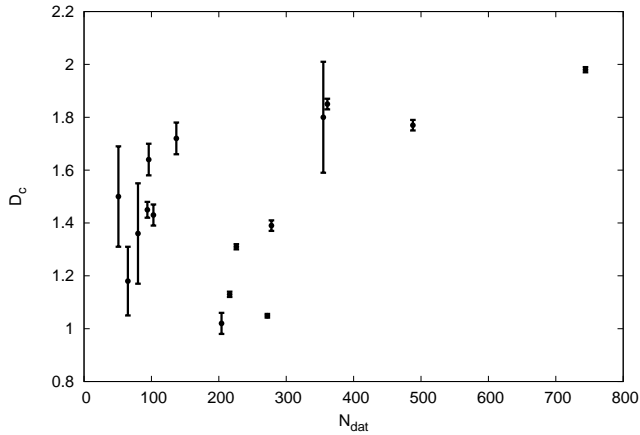} 
\caption{Fractal (correlation) dimension for the
distribution oy young stars and pre-main-sequence
stars in different star-forming regions as a function
of the number of data points (Table~\ref{tabDc}).}
\label{figDc}
\end{center}
\end{figure}

The fractal dimension of a distribution of stars can be measured
by using the correlation integral $C(r)$.
For a fractal set it holds that
$C(r) \sim r^{D_c}$, being $D_c$ the so-called correlation
dimension. Calculating the mean surface density of companions
(MSDC) per star $\Sigma(\theta)$ as a function of angular
separation $\theta$ is another widely used way to measure
the degree of clustering of stars. For fractals $\Sigma(\theta) 
\sim \theta^\gamma$, and the exponent is related to the fractal
dimension through $D_c=2+\gamma$.
MSDC technique has been used by various authors to study the 
clustering of protostars, pre-main sequence stars, or young 
stars in different star-forming regions. Most results seem to 
indicate that there are two different ranges of spatial scales, 
the regime of binary and multiple systems on smaller scales and 
a regime of fractal clustering on the largest scales. The idea 
prevalent among astronomers is that self-similar clustering above 
the binary regime is due to, or arises from, the fractal features 
of the parent clouds. However, such as in the 
case of gas distribution in the ISM, if one checks the references a 
wide variety of different values can be found. Nakajima et al. 
(1998) found significantly variations among different star-forming 
regions with $1.2 \lesssim D_c \lesssim 1.9$. There can be large
differences even in the same regions if analyzed by different
authors and data sets. For example, in the Taurus region both
Larson (1995) and Simon (1997) analyzed young stars and their
results are in perfect agreement with $D_c \simeq 1.4$, whereas
Hartmann (2002) and Kraus \& Hillenbrand (2008) both agree in
$D_c \simeq 1.0$ for the same region. 
Table~\ref{tabDc} summarizes the wide range of $D_c$ values 
estimated by using the MSDC technique.
Obviously, there can be different results depending 
on data sources, object selection criteria, and details of 
the specific calculation procedures. But additionally, it has 
been shown that if boundary and/or small data-set effects 
are not taken into account the final results can be seriously 
biased, given fractal dimension values smaller than the true 
ones (S\'anchez et al. 2007b; see also S\'anchez \& Alfaro 2008). 
In Figure~\ref{figDc} we have plotted $D_c$ values from
Table~\ref{tabDc} as a function of the number of data points
$N_{dat}$. The observed behavior seems to be biased (at least
in part) in the sense that $D_c$ decreases as $N_{dat}$ decreases
(compare Figure~\ref{figDc} here with Figures~$2$ and $4$ in
S\'anchez \& Alfaro 2008). If this kind of effect is not corrected
then any real variation in $D_c$ could be hidden or misunderstood.
\begin{table}
\begin{center}
\caption{Summary of correlation dimensions for the distribution 
of stars in clusters}
\label{tabDc}
{\scriptsize
\begin{tabular}{|c|c|l|l|}\hline 
Ref. & $N_{dat}$ & $D_c$ Ê& Cluster \\
\hline
(1) & $>121$ & $1.38$          & Taurus-Auriga \\
(2) & $80$   & $1.36\pm0.19$   & Taurus \\
    & $51$   & $1.50\pm0.19$   & Ophiuchus \\
    & $355$  & $1.80\pm0.21$   & Trapezium \\
(3) & $361$  & $1.85\pm0.02$   & Orion OB \\
    & $488$  & $1.77\pm0.02$   & Orion A \\
    & $226$  & $1.31\pm0.01$   & Orion B \\
    & $96$   & $1.64\pm0.06$   & Ophiuchus \\
    & $103$  & $1.43\pm0.04$   & Chamaeleon I \\
    & $94$   & $1.45\pm0.03$   & Chamaeleon \\
    & $278$  & $1.39\pm0.02$   & Vela \\
    & $65$   & $1.18\pm0.13$   & Lupus \\
(4) & $744$  & $1.98\pm0.01$   & Trapezium \\
(5) & $137$  & $1.72\pm0.06$   & Chamaeleon I \\
    & $216$  & $1.13\pm0.01$   & Taurus \\
(6) & $204$  & $1.02\pm0.04$   & Taurus \\
(7) & $272$  & $1.049\pm0.007$ & Taurus-Auriga \\
\hline
\end{tabular}}
\end{center}
\vspace{1mm}
\scriptsize{
{\it Reference index:}
(1) Larson (1995); 
(2) Simon (1997); 
(3) Nakajima et al. (1998); 
(4) Bate et al. (1998), their first data set; 
(5) Gladwin et al. (1999); 
(6) Hartmann (2002); 
(7) Kraus \& Hillenbrand (2008).}
\end{table}

\section{Evolutionary effects}

Even in the case of extremely young cluster, we always are seeing 
a snapshot of the cluster at a particular age resulting from certain 
initial conditions (ISM structure) and early dynamical evolution. As 
a cluster evolves, its initial distribution of stars may be erased, 
or at least modified. Then, part of the variations observed in
Figure~\ref{figDc} could be due to evolutionary effects.
The early evolution of the cluster will depend, among other things,
on how much gas is removed after the formation process (Gieles 2010).
In gravitationally unbound clusters, the separation of the stars
increases with age until the cluster dissolves into the field.
In principle, the initial clumpy structure disappears after
this process of expansion although some simulations suggest
that it is possible to keep the initial substructure for a
long time in unbound clusters (Goodwin \& Whitworth 2004).
Gravitationally bound clusters, instead, have to evolve toward
a new equilibrium state. 
Simulations show that this dynamical evolution can be a very complex 
process (e.g., Moeckel \& Bate 2010). It seems that the general trend 
is to evolve from the initially substructured distribution of stars 
toward centrally peaked distributions, that is, radial star density 
profiles. The evidence for this kind of evolution comes from both 
observations and from numerical simulations (Schmeja \& Klessen
2006; Schmeja et al. 2008; S\'anchez \& Alfaro 2009; Allison et 
al. 2009, 2010; Moeckel \& Bate 2010).

Roughly speaking, the time interval necessary to erase any initial 
structure will depend on the crossing time $T_{cross}$.
It should take at least several crossing times to reach an 
equilibrium state and/or to eliminate the original distribution,
although some simulations indicate 
that the evolution from clumpy to radial distribution may occur on 
time scales as short as $\sim 1$ Myr (Allison et al. 2009, 2010).
In order to address these questions, it is necessary to characterize 
the internal structure of young clusters and also to get some idea 
about the evolutionary stage.

\section{Minimum spanning tree}

For radially concentrated clusters, star distribution ca be
characterized by fitting the density profile to some given
predefined function. From the fitting procedure it is possible
to get parameters such as the central density of stars, the
steepness of the density profile, and cluster radius.
Obviously, this kind of analysis does not work in clumpy
clusters because a smooth function cannot be well fitted to
an irregular distribution.

\begin{figure}[thb]
\begin{center}
\includegraphics[width=\hsize]{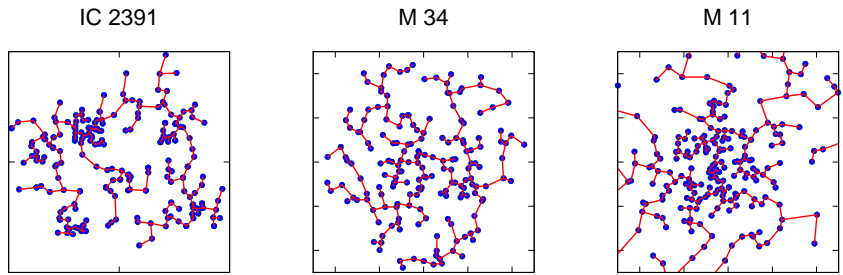} 
\caption{Minimum spanning trees for three open clusters, from which
the parameter $Q$ can be calculated (see text). Star positions are
indicated with blue circles and red lines represent the trees. The
value of $Q$ quantifies the spatial distribution of stars. For
IC~2391 the stars are distributed following an irregular, fractal
pattern ($Q = 0.66 < 0.8$), for M~34 the stars are distributed
roughly homogeneously ($Q=0.8$), and for M~11 the stars follow
a radial density profile ($Q = 1.02 > 0.8$).}
\label{figMST}
\end{center}
\end{figure}

Cartwright \& Whitworth (2004) proposed a method to 
quantify the internal structure of star clusters. Their technique 
is becoming very widespread and useful for analyzing both 
observational and simulated data because, it is able to 
distinguish between centrally concentrated and fractal-like 
distributions. The technique is based on the construction of 
the minimum spanning tree (MST). The MST is the set of straight 
lines (called branches or edges) connecting a given set of points 
without closed loops, such that the total edge length is minimum 
(see Figure~\ref{figMST}). From the MST an adimensional structure
parameter $Q$ can be easily calculated (Cartwright \& Whitworth
2004; see also Schmeja \& Klessen 2006). For an homogeneous
distribution of stars $Q \simeq 0.8$. The behavior of $Q$ is
such that $Q > 0.8$ for radial clustering whereas $Q < 0.8$ for
fractal clustering (see the examples in Figure~\ref{figMST}).
Moreover, $Q$ increases as the steepness
of the profile increases for radial clustering and $Q$ decreases
as the fractal dimension decreases for fractal-type clustering
(see Figure~$5$ in Cartwright \& Whitworth 2004; and Figure~$7$
in S\'anchez \& Alfaro 2009). Thus, $Q$ is able to disentangle
between radial and fractal clustering but it also measure the
strength of clustering.

\begin{figure}[thb]
\begin{center}
\includegraphics[width=\hsize]{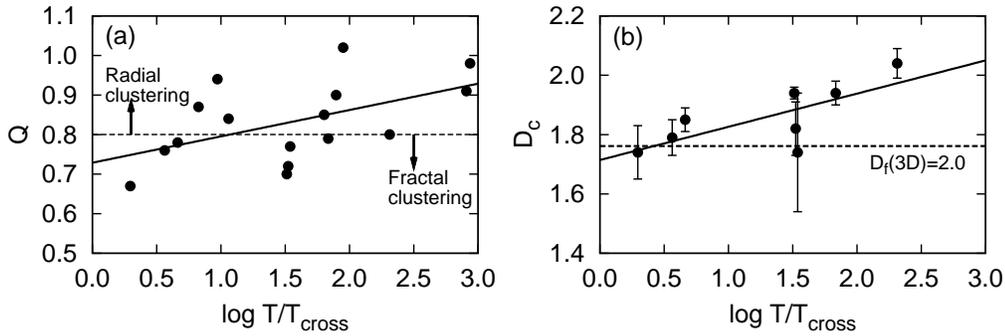}
\caption{(a) Structure parameter as a function of the
logarithm of age in crossing time units. Dashed line
separates radial from fractal clustering and
solid line is the best linear fit.
(b) Correlation dimension as a function of age in
crossing-time units. The best linear fit is represented 
by a solid line. For reference, a horizontal dashed line
indicates the value corresponding to three-dimensional
distributions with fractal dimensions of $D_f = 2.0$.}
\label{figcorrela}
\end{center}
\end{figure}

It is expected that the internal structure of a star 
cluster evolves with time from initial fractal clustering ($Q < 0.8$) 
to either homogeneous distribution ($Q \simeq 0.8$) if the cluster is 
dispersing its stars or centrally concentrated distribution 
($Q > 0.8$) if it is a bound cluster. Cartwright \& Whitworth 
(2004) calculated $Q$ FOR several star clusters. 
They obtained $Q=0.47$ for stars in Taurus, a value consistent 
with its observed clumpy structure and with its relatively
young evolutionary stage (they estimated an age in crossing
time units of $T/T_{cross} \simeq 0.1$). However, they 
also found some apparent contradictory results. IC~348, a 
slightly evolved cluster with $T/T_{cross} \simeq 1$ yielded
$Q=0.98$ according to its steep radial density profile.
Instead, the highly evolved cluster IC~2391 ($T/T_{cross}
\simeq 20$) still exhibits fractal clustering with $Q=0.66$.
Schmeja et al. (2008) applied this technique to
embedded clusters in the Perseus, Serpens and 
Ophiuchus molecular clouds, and found that older Class 2/3 
objects are more centrally condensed than the younger Class 
0/1 protostars. S\'anchez \& Alfaro (2009) measured $Q$ in a 
sample of $16$ open clusters in the Milky Way
spanning a wide range of ages. 
They found that there can exist clusters as old as $\sim 100$
Myr exhibiting fractal structure.
This means that either the initial clumpiness may 
last for a long time or other mechanisms may develop some 
kind of substructure starting from an initially more 
homogeneous state. S\'anchez \& Alfaro (2009) obtained 
a statistically significant correlation between $Q$ and 
$T/T_{cross}$ in their sample of open clusters. 
Figure~\ref{figcorrela}a shows
this tendency where the crossing times were 
calculated by assuming a constant velocity dispersion 
of $2$ km s$^{-1}$. As we can see, the general trend is
that young clusters (meaning that dynamically less evolved 
clusters) tend to distribute their stars following 
fractal patterns whereas older clusters tend to 
exhibit centrally concentrated structures. This 
result support the idea that stars in newly born 
clusters likely follow the fractal patterns of their 
parent molecular clouds, and that they eventually 
evolve towards more centrally concentrated structures. 
However, we know that this is only an overall trend. 
Some very young clusters may exhibit radial density 
gradients, as for instance $\sigma$~Orionis for which
$Q \simeq 0.9$ (Caballero 2008).

Given the wide variety of physical processes involved 
in the origin and early evolution of star clusters, it 
is somewhat surprising that a correlation like that seen 
in Figure~\ref{figcorrela}a can be observed. Very recent
simulations by Allison et al. (2009, 2010) and Moeckel \&
Bate (2010) show that the transition from fractal clustering
to central clustering may occur on very short timescales
($\lesssim 1$ Myr). Simulations by Maschberger et al. (2010)
suggest a more complex variety of possibilities. Bound systems
may start fractal and evolve towards a centrally concentrated
stage whereas unbound systems may stay fractal in time. But
this is the evolution for the whole systems. Star clusters
in each system may evolve in totally different ways. In fact,
the time evolution of the $Q$ parameter of clusters fluctuates
dramatically depending on episodes of relaxation or 
merging (see Figure~$8$ in Maschberger et al. 2010).
It is difficult to argue that, despite all this complex
formation history (occurring in $\sim 0.5$ Myr), we
should still observe some correlation between
internal structure and age.

\section{Initial fractal dimension of star clusters}

For those open clusters with internal substructure, the
fractal dimension also shows a significant correlation with
the age in crossing time units (S\'anchez \& Alfaro 2009), as
it can be seen in Figure~\ref{figcorrela}b.
The degree of clumpiness is smaller for more evolved
clusters. The horizontal dashed line in
Figure~\ref{figcorrela}b shows a reference $D_f$
value estimated from previous papers (S\'anchez \&
Alfaro 2008). It is interesting to note that open
clusters with the smallest correlation dimensions
$D_c = 1.74$ would have 3D fractal dimensions around
$D_f \sim 2$. This value is considerably smaller
than the average value estimated for Galactic molecular 
clouds (see Section~$2$), which is $D_f \simeq 2.6-2.8$.

This result creates an apparent problem to be
addressed, because as mentioned before 
a group of stars born from the same cloud at almost the 
same place and time is expected to have a fractal dimension 
similar to that of the parent cloud. If the fractal dimension 
of the interstellar medium has a nearly universal value around 
$2.6-2.8$, then how can some clusters exhibit such small fractal 
dimensions? This is still an open question. Several possibilities 
should be investigated in future studies. First, some simulations 
demonstrate that it is possible to increase the clumpiness (to 
decrease $D_f$) with time (Goodwin \& Whitworth 2004). Second,
maybe this difference is a consequence of a more clustered
distribution of the densest gas from which stars form on the
smallest spatial scales in the molecular cloud complexes,
according to a multifractal scenario (Chappell \& Scalo 2001).
Third, perhaps the star formation process itself modifies in some 
(unknown) way the underlying geometry generating distributions 
of stars that can be very different from the distribution of 
gas in the star-forming cloud. A fourth possibility is that
the fractal dimension of the interstellar medium in the
Galaxy does not have a universal value (i.e., that $D_f$
is different from region to region depending on
the main physical processes driving the turbulence).
Therefore some clusters could show smaller initial fractal
dimensions because they formed in more clustered regions.
The possibility of a non-universal fractal dimension for the ISM
should not, in principle, be ruled out. However, in this last case,
overall correlations as those shown in Figures~\ref{figcorrela}a
and \ref{figcorrela}b should not, in principle, be observed.

\end{document}